\documentclass{ws-ijmpe}
\usepackage[utf8]{inputenc}
\usepackage{graphicx}
\usepackage[numbers,compress]{natbib}
\usepackage{multicol}
\usepackage{amsmath}
\usepackage{slashed}
\usepackage{commath}
\usepackage{amssymb}
\usepackage{enumerate}
\usepackage{physics}
\usepackage{comment}
\usepackage{xcolor}

\usepackage[colorlinks]{hyperref}
\begin{document}

\catchline{}{}{}{}{}
\title{Nuclear response to dark matter signals in Ge and Xe odd-mass targets}
\author{M. M. Saez}
\address{Interdisciplinary Theoretical and Mathematical Sciences Program (iTHEMS), RIKEN, Wako, Saitama 351-0198, Japan and Department of Physics, University of California, Berkeley, CA 94720\\
manuelasz89@gmail.com}

\author{O. Civitarese\footnote{corresponding author}}
\address{Dept. of Physics, University of La Plata, c.c.~67, (1900) La Plata, Argentina\\
Institute of Physics. IFLP-CONICET. diag 113 e/63-64.(1900) La Plata. Argentina\\
osvaldo.civitares@fisica.unlp.edu.ar}

\author{T.Tarutina}
\address{Dept. of Physics, University of La Plata, c.c.~67, (1900) La Plata, Argentina\\
Institute of Physics. IFLP-CONICET. diag 113 e/63-64.(1900) La Plata. Argentina\\
tatiana.tarutina@fisica.unlp.edu.ar}

\author{K. Fushimi} 
\address{Faculty of Astronomy and Geophysics,University of La Plata.\\
Ps del Bosque s/n (1900) La Plata, Argentina\\
keiko.fushimi89@gmail.com}

\date{\today}
\maketitle

\begin{history}
\received{Day Month Year}
\revised{Day Month Year}
\end{history}

\begin{abstract}
Abstract: The interaction of dark matter particles (WIMPs) with the odd-mass $^{73}$Ge and $^{131}$Xe target nuclei 
,{  {that is the recoil rates corresponding to the elastic scattering of WIMPs by these nuclei}},
is analysed in the context of the minimal extensions of the SUSY model. The BCS+QRPA technique plus the quasiparticle-phonon coupling scheme  is used to describe the nuclear structure part of the calculations. The resulting values for the nuclear spin content of both nuclei are compared to values previously reported in the literature.
\end{abstract}


\ccode{PACS numbers: }


\section{Introduction}

The nature of dark matter (DM) remains one of the most pressing issues in modern physics. Dark 
matter has not been detected directly yet. Still, there is evidence extracted from observations at galactic scales, galaxy clusters and cosmological observables which suggests that much of the Universe is dark \cite{Rubin:1970,zwicky:1933,Clowe:2006,Planck:2018,Schumann:2019}. 
A well-established paradigm is that most DM is cold and is made up of weakly interacting massive particles (WIMPs); other promising alternatives are axions \cite{majumdar:2014,Peccei:1977,chadhaday:2021}.
The WIMPs as cold dark matter candidates arise naturally in various theories beyond the Standard Model, e.g., the lightest supersymmetric particle is the neutralino $\chi$, with an expected mass between $1-1000$~GeV \cite{Gelmini:2017}.A relevant strategy for searching WIMPs is the direct detection through {  {
the elastic scattering of DM particles by nuclei in ultrasensitive low background experiments}}\cite{Goodman:1985, Wasserman:1986}. 
WIMPs interact with ordinary matter predominantly through axial-vector (spin-dependent) and scalar (spin-independent) couplings.
Given that the { {elastic scattering cross-section}} for the spin independent part is proportional to the square of the nuclear mass number, it is convenient to use heavy nuclei as targets,  but to be sensitive to the spin-dependent interaction, a target with an odd number of protons or neutrons must be used. 

Some leading dark matter experiments use liquid xenon or germanium as  targets \cite{XENON1T:2018,LUX:2017,EDELWEISS:2020,CoGeNT:2012}.

In this work, we focus on the spin-dependent (SD) interaction of WIMPs with the odd mass nuclei $^{73}{\rm Ge}$ and $^{131}{\rm Xe}$,{  {to calculate the differential recoil rates for the elastic scattering of WIMPs by these nuclei
}}. The SD interaction of DM particles can constrain WIMPs properties, giving more substantial limits on the SUSY parameter space than the spin-independent interaction and it is particularly sensitive to the nuclear structure of the involved nuclei \cite{Bednyakov:2007,Menendez:2012}.

{ {The particular choice of $^{73}{\rm Ge}$ and $^{131}{\rm Xe}$ is motivated by both theoretical and experimental reasons. 
Previous calculations of cross-sections and reaction rates for SD interactions have been performed at zero momentum transfer \cite{Marcos:2016,Bednyakov:2004} but recent works have shown that the dependence of the nuclear matrix elements on the momentum transfer cannot always be ignored \cite{Bednyakov:2007,Menendez:2012,Bednyakov:2006,Vietze:2014}. In the present case,we are applying the Quasiparticle Random Phase Approximation (QRPA) to describe monopole and quadrupole low-lying excitations of the even mass cores  $^{72}{\rm Ge}$ and $^{130}{\rm Xe}$, and coupled a quasi-neutron to them. The results obtained by using the quasiparticle-phonon model\cite{Bohr-Mottelson:1974} are going to be compared to the results obtained by other authors in the context of the Interacting Boson Model for odd mass nuclei \cite{Pirinen:2019}, the Tamm-Dancoff approach \cite{Engel:1992}) and the Chiral Effective Field Theory \cite{Klos:2013}.
From the experimental side, the planned measurements of the XENON1T Collaboration \cite{XENON1T:2018} for Xe isotopes, and 
 LUX\cite{LUX:2017}, EDELWEISS \cite{EDELWEISS:2020} and CoGeNT\cite{CoGeNT:2012}. for Ge isotopes could provided relevant information about the elastic scattering of WIMPs by nuclei}}.

The paper is organized as follows. In Section \ref{formalismo_wimps} we present a brief description of 
the dark matter model and the formalism needed to compute direct detection rates, cross-sections and form factors both for zero and finite momentum transfers. In Section \ref{nuclear} we focus on the properties of the target nuclei and their microscopic description using the quasiparticle-phonon coupling model \cite{Bohr-Mottelson:1974}. In Section \ref{results} we present  and discuss the results of the calculations and compare our results with those of other works \cite{Bednyakov:2007,Engel:1992, Pirinen:2019}. We shall follow, as closely as possible, the expressions presented in Refs.\cite{Engel:1992, Pirinen:2019}, because of the relevance of both for the purpose of the comparison between the results given by the TDA (Engel et al.\cite{Engel:1992}), IBM (Pirinen et al. \cite{Pirinen:2019}) and QRPA (present work) models, for the nuclear structure part of the calculations.

The conclusions are drawn in Section \ref{conclusions}.

\section{Dark matter direct detection}\label{formalismo_wimps}
\subsection{Dark matter model}\label{dmmodel}
We work within the framework of the MSSM (Minimal Supersymmetric Standard Model), in which the lightest neutralino state can be written as a linear combination of binos, winos, and higgsinos ($\tilde{B}$, $\tilde{W_3}$, $\tilde{H}_1$ and $\tilde{H}_2$) 

\begin{equation}
      \chi_1^0=Z_{11}\tilde{B}+Z_{12}\tilde{W_3}+Z_{13}\tilde{H}_1+Z_{14}\tilde{H}_2\,.
      \label{chi}
 \end{equation}
The coefficients of the linear combination $(Z_{11},\, Z_{12},\, Z_{13},\, Z_{14})$ depend on four SUSY parameters, the bino and wino mass parameters ($M'$ and $M$), the higgsino mass parameter ($\mu$), and the value of $\tan{\beta}$ related to the ratio of vacuum expectation values of the two Higgs scalars. In the Grand-Unified-Theory (GUT), the parameters $M$ and $M'$ are related by $M' = \frac{5}{3} M \tan^2{\theta_W}$ \cite{Murakami:2001,Ellis:2000,Cerdeno:2001}.
The neutralino-quark elastic scattering Lagrangian density in the MSSM is written as \cite{Engel:1992}
\begin{equation}
  \mathcal{L}_{eff}=\frac{g^2}{2M_W^2}\sum_q\left(\bar{\chi}\gamma^{\mu}\gamma_5\chi\bar{\psi_q}\gamma_{\mu}A_q\gamma_5\psi_q+\bar{\chi}\chi S_q \frac{m_q}{M_W}\bar{\psi_q}\psi_q\right) ,
\label{lagrangiano}
\end{equation}
where $M_W$ stands for the mass of the W boson, $g$ is the SU(2) coupling constant, and $A_q$ and $ S_q$ are defined by \cite{Engel:1992}
\begin{eqnarray}
  A_q&=&\frac{1}{2} T_{3q}(Z^2_{13}-Z^2_{14})\nonumber\\
  &&-\frac{M^2_W}{M^2_{\tilde{q}}}\Bigg(\left[T_{3q}Z_{12}-(T_{3q}-e_q)Z_{11}\tan{\theta_W}\right]^2\nonumber\\
  &&\hspace{1.4cm}+e_q^2Z_{11}^2\tan^2{\theta_W}+\frac{2m_q^2d_q^2}{4M_W^2}\Bigg)\,,
  \label{A_q}
\end{eqnarray}
\begin{eqnarray}
S_q&=&\frac{1}{2}(Z_{12}-Z_{11}\tan{\theta_W})\Bigg[\frac{M^2_W}{M^2_{H_2}}g_{H_2}k_q^{(2)}+\frac{M^2_W}{M^2_{H_1}}g_{H_1}k_q^{(1)}\nonumber\\
&&\hspace{3.5cm}+\frac{M^2_W \epsilon d_q}{M^2_{\tilde{q}}}\Bigg]\,,
\end{eqnarray}
where $M_{\tilde{q}}$ and $M_{H_i}$ are the squark and higgsino
masses respectively \cite{Djouadi:2008,Pdg:2019}, $T_{3q}$,
$e_q$, and $m_q$ are the quark weak isospin, charge and mass
respectively, and $\epsilon$ is the sign of the lightest-neutralino mass
eigenvalue \cite{Engel:1992}. The $d_q$ and $k_q^{(i)}$ parameters for the up and down quarks were taken from  \cite{Djouadi:2008}. We compute the cross-section using the Lagrangian of Eq.~\eqref{lagrangiano}. See Ref. \cite{Fushimi:2020} and references therein for more details about the calculations.

\subsection{Detection rates:elastic scattering of WIMPs by nuclei}\label{sec:rates}

The differential recoil rate per unit mass of the detector can be defined as \cite{Freese:2012}
\begin{eqnarray}\label{rate2}
\frac{dR}{dE_{\rm nr}}&=&\frac{2\rho_{\chi}}{m_{\chi}} \frac{\sigma_0}{4 \mu^2 } F^2(q) \eta\, ,
\end{eqnarray}
in units of $\textrm{cpd}\,\textrm{kg}^{-1} \textrm{keV}^{-1}$, where cpd stands for counts per day. In the previous equation $E_{\rm nr}$ is the nuclear recoil energy,  $q^2= 2 m_A E_{\rm nr}$, $m_A$ is the mass of the nucleus, $m_{\chi}$ and $\rho_{\chi}$ are the WIMP mass and  local density, $\rho_{\chi}=0.3$~GeV/cm$^3$ \cite{Schumann:2019,Gelmini:2017}, $\sigma_0$ is the cross-section at $q=0$ and $F(q)$ stands for the nuclear form factor. We define the mean inverse-velocity as
\begin{eqnarray}
\label{inv-vel}
\eta&=&\int{\frac{f(\vec{v},t)}{v}d^3v} \, ,
\end{eqnarray}
$f(\vec{v},t)$ is the WIMP velocity distribution and $v$ is the speed of the WIMP relative to the nucleus.

\subsubsection{Velocity distribution}

For the velocity distribution of the WIMP, we assume that the model for the DM halo is the Standard Halo Model \cite{Freese:1988}; therefore, we calculate the velocity distribution from the truncated Maxwell-Boltzmann distribution \cite{Freese:2012,Jungman:1996}
\begin{eqnarray}\
f({\vec{v}})&=&\left\{
\begin{array}{cc}
\frac{1}{N(\pi v_0^2)^{\frac{3}{2}}}e^{{-\left|{\vec{v}}\right|^2}/{v_0^2}} & |\vec{v}|<v_{\rm esc} \\
0&|\vec{v}|>v_{\rm esc} \\
\end{array}
\right. \, , 
\end{eqnarray}
where $N$ is a normalization factor 
and $v_{\rm esc}$ and $v_0$ are the escape velocity and the velocity of the Sun, respectively, and their values are $v_{\rm esc}=544 \, {\rm km/s}$ \cite{Gelmini:2015} and $v_0=220 \, {\rm km/s}$ \cite{Jungman:1996}. If one considers the laboratory-velocity $\left(\vec{v}_{\rm lab}\right)$, the integral of Eq.~\eqref{inv-vel} becomes

\begin{equation}\label{eta1-2}
\eta=\frac{1}{N{(\pi {v_0}^2)}^{3/2}}\int \frac{e^{-{(\vec{v}+\vec{v}_{\rm lab})}^2/{v_0^2}}}{v}d^3v\, .
\end{equation}

\subsubsection{Elastic scattering cross-section at zero momentum transfer}
The total WIMP-nucleus { {elastic scattering cross-section}} is a sum over the spin-independent (SI) and spin-dependent (SD) contributions. Since, in this work, we are dealing with the spin dependent channel, the differential cross-section , 
{ {for the elastic scattering process}}
is written \cite{Freese:2012}:
\begin{equation}\label{dif-seccioneficaz-nucleo}
\frac{d\sigma}{dE_{\rm nr}}(E_{\rm nr},v)= \frac{m_A}{2v^2\mu_{A}^2}\sigma_{SD}(0)F_{SD}^2(q)\,,
\end{equation}
where $\mu_{A}$ is the reduced mass of the WIMP-nucleus system. For this cross-section we consider all the mediators that contribute to the axial-vector interaction, that is $ Z$ and $ \tilde{q} $.

In the zero momentum transfer approximation, we assume that the nuclear form factor $F_{SD}$ is a constant \cite{Marcos:2016}. The cross-section  $\sigma_{SD}(0)$ is proportional to the total angular momentum of the nucleus $J$ and it could be written as a function of the neutralino-proton cross-section ($\sigma_{SD}^p$) \cite{Marcos:2016}

\begin{equation}\label{sigma_SD}
\sigma_{SD}(0)=\sigma_{SD}^p\left[\langle S_p \rangle+ \langle S_n \rangle\frac{a_n}{a_p}\right]^2\left(\frac{\mu_{A}}{\mu_{p}}\right)^2\frac{4 (J+1)}{3J}\,.
\end{equation}

where $\mu_{p}$ is the reduced mass of the WIMP-proton system and $\langle S_{n(p)}\rangle$ is the spin expectation value of the nucleon over the nuclear wave function, the coefficients  ${a_{n,(p)}}$ are the strengths of the coupling of the WIMPs to the nucleons \cite{Engel:1992}.

\subsubsection{Elastic scattering cross-section at finite momentum transfer}
The cross-section for this case is also given by Eq. \ref{dif-seccioneficaz-nucleo} with the form factor 
 
\begin{equation}
F_{SD}^2(q)= \frac{S_{SD}(q)}{S_{SD}(0)}.
\end{equation}
The axial structure function $S_{SD}(q)$ can be expressed as \cite{Engel:1992}
\begin{equation}
S_{SD}(q)=a_0^2S_{00}(q)+a_1^2S_{11}(q)+a_0a_1S_{01}(q)\,.
\end{equation}
Some results about the structure functions  $S_{ij}$, for the case of $^{73}$Ge and $^{131}$Xe, are given in Refs. \cite{Bednyakov:2006,Sahu:2016}. They are 
extracted from calculations based on nuclear models, as we shall see  next, in the context of the quasiparticle-phonon coupling model. The constants $a_0$ and $a_1$ are related to $a_n$ and $a_p$ via \cite{Engel:1992}
\begin{align}
a_0=a_n+a_p,\nonumber\\
a_1=a_p-a_n
\end{align}
where $a_n$ and $a_p$ are expressed in terms of their up and down quark content \cite{Weinberg}.


\section{Nuclear structure aspects of the calculations}\label{nuclear}

To compute the spin expectation values which appear in Eq.(\ref{sigma_SD}) we have applied the quasiparticle-phonon coupling model \cite{Bohr-Mottelson:1974}. For the results of calculations performed in the context of other models we shall refer
and compare our results with those of Refs.\cite{Menendez:2012},\cite{Engel:1992} and \cite{Pirinen:2019}, as well as those obtained  using average spin values \cite{Tovey:2000,Dimitrov:1994}. 
Both $^{73}$Ge and $^{131}$Xe odd-mass targets are nuclei with active neutron excess.
{ {
 As explained in the Introduction both nuclei are or should be the  materials of choice for current and future experimental efforts \cite{XENON1T:2018, LUX:2017,EDELWEISS:2020,CoGeNT:2012}.}}
 
To describe their microscopic structure we shall proceed by applying a coupling scheme of neutrons with the low-lying excitations of the even mass $^{72}$Ge and 
$^{130}$Xe nuclei. The method is rather well known and it is a matter of textbooks, among them \cite{Bohr-Mottelson:1974} and \cite{Ring-Schuck:1980}, so we shall briefly outline the main steps of the theoretical framework, which have been presented also in some of our previous works \cite{Civitarese:1998}. They are the following:

\begin{itemize}
\item{Use of the BCS transformations to the quasiparticle basis}
For neutron states in the single particle states belonging to the 28 $\leq$ N  $\leq$ 50 and 50 $\leq$ N $\leq$ 82  neutron orbits in the central harmonic oscillator potential with parameters taken from Ref.\cite{Bohr-Mottelson:1974}, for Ge and Xe isotopes, respectively, the value of the pairing gap of the even mass  $^{72}$Ge and $^{130}$Xe nuclei have been determined by adjusting the pairing coupling constants to $g=20/A$ MeV of the monopole pairing interaction for each of them.

\item{Use of the Quasiparticle Random Phase Approximation (QRPA)}

The well known QRPA method \cite{Bohr-Mottelson:1974} is use to diagonalize two- and four-quasiparticle terms of separable monopole and quadrupole interactions of the type 
\begin{eqnarray}
H_{(\lambda)}&=&\sum_{a}E_{a} A^{\dagger}(a,\lambda \mu)A(a,{\overline{\lambda \mu}})\nonumber \\
         &+&\sum_{a,b}F_{ab:\lambda}A^{\dagger}(a,\lambda \mu)A(b,{\overline{\lambda \mu}})\nonumber \\
         &+&\sum_{a,b} G_{ab:\lambda}(A^{\dagger}(a,\lambda \mu)A^{\dagger}(b,{\overline{\lambda \mu}})
        +A(a,\lambda \mu)A(b,{\overline{\lambda \mu}})),
\end{eqnarray}
where $A^{\dagger}(a,\lambda \mu)$ and $A(a,{\overline{\lambda \mu}})$ denote the creation and annihilation of two-quasiparticle configurations of energy E$_a$ coupled to total angular momentum ${\lambda}$ and its projection $\mu$, and where $F_{ab:\lambda}$ and $G_{ab:\lambda}$ are the matrix elements of the separable multipole interaction 
\begin{equation}
H_{int}=\sum_{\lambda \mu} \chi_{\lambda} Q^{\dagger}_{{\lambda \mu}}Q_{\overline{\lambda \mu}},
\end{equation}
written in the quasiparticle basis,
being $Q^{\dagger}_{{\lambda \mu}}$ and $Q_{\overline{\lambda \mu}}$ the monopole ($\lambda =0$) and quadrupole
($\lambda=2$) creation and annihilation operators, respectively. The coupling constant $\chi_{\lambda}$, for each channel, is adjusted in order to reproduce the energy $\hbar w_\lambda$ of the first excited monopole and quadrupole states of the even mass 
$^{72}$Ge and $^{130}$Xe nuclei
\begin{equation}
\frac{1}{\chi_{\lambda}}=\sum_a\frac{2E_a q_a^2(\lambda)}{E_a^2-\hbar w_\lambda^2},
\end{equation}
where the reduced matrix elements of the multipole operator in the quasiparticle basis, $q_a(\lambda)$ for pairs $a=(1,2)$ of quasiparticles coupled to $(\lambda,\mu)$ are written
\begin{equation}
q_a(\lambda)=(u_1v_2-(-1)^cv_1u_2)\frac{\langle 1 ||Q_\lambda|| 2 \rangle}{\sqrt{2.\lambda+1}},
\end{equation}
$u_i$ and $v_i$ are the quasiparticle occupation factors and $(-1)^c$ is the phase determined by time reversal for 
multipole operators \cite{Bohr-Mottelson:1974}.
The transformation to the monopole and quadrupole one phonon states yields 
\begin{equation}
H_{phonon}=\sum_{\lambda \mu} \hbar w_\lambda \Gamma_{(\lambda \mu)}^\dagger \Gamma_{\overline{\lambda \mu}}+constant,
\end{equation}
where $\Gamma_{(\lambda \mu)}^\dagger$ and $ \Gamma_{(\lambda \mu)}$ are one phonon creation and annihilation operators which are written
\begin{eqnarray}
\Gamma_{(\lambda \mu)}^\dagger &=&\sum_a( X_a A^{\dagger}(a,\lambda \mu)-Y_a A(a,\overline{\lambda \mu}))\nonumber \\
\Gamma_{(\lambda \mu)}&=&\sum_a (X_a A(a,\lambda \mu)-Y_a A^{\dagger}(a,\overline{\lambda \mu})),
\end{eqnarray}
with forward and backward going amplitudes $X_a$ and $Y_a$, respectively.
 
\item{Diagonalization of the one-quasiparticle plus one phonon couplings}

The last step towards the determination of the wave functions and energies of low-lying states belonging to the odd mass Ge and Xe isotopes consists of transforming the remaining terms of the Hamiltonian, the so-called $H_{31}$ terms of it \cite{odd}, to the one quasiparticle plus one phonon basis, that is by diagonalizing the interaction

\begin{equation}\label{hodd}
H_{qp-ph}=\sum_{\lambda} \Lambda_{\lambda} {(Q_{\lambda}(\Gamma_{\lambda}^\dagger + \Gamma_{\lambda}))}_0,
\end{equation}

where the quantities  $\Lambda_{\lambda}$  are the strength of the couplings whose expression in terms of the QRPA amplitudes and matrix elements of the multipole operators is given by 

\begin{equation}
\Lambda_{\lambda}=-\chi_{\lambda}\sum_a(X_a+Y_a)q_a(\lambda).
\end{equation}

\item{Wave functions of the odd-mass nuclei}
 
The diagonalization of the term $H_{qp-ph}$ of Eq.(\ref{hodd}) in the basis of one quasiparticle coupled to n (n=0,1) phonon states of multipolarity $\lambda$ gives the wave functions
\begin{equation}
    |IM\rangle = \sum_{j,n=0,1} \alpha(j,n;I)|j \otimes \lambda ;IM\rangle,
\end{equation}
being $\alpha(j,n;I)$ the corresponding amplitudes.
\end{itemize}

The nuclear response to the interaction with dark matter particles, as described by the cross section of Eq.(\ref{sigma_SD}),
is given by the expectation value of the spin channel taken on the wave function of the states belonging to the spectrum of the odd-mass nuclei, which is written

\begin{equation}
\langle IM|S_{1\mu}|IM\rangle = \frac{1}{\sqrt{2I+1}} \langle IM1\mu|IM\rangle  \langle I||S_{1}||I\rangle,
\end{equation}

where

\begin{eqnarray} \label{matrix_elements}
\langle I||S_{1}||I\rangle&=&
    (2I+1) \sum_{j,n_\lambda, j', n_\lambda'} \alpha(j,n_\lambda;I) \alpha(j',n_\lambda';I) (-1)^{l+1+1/2+I}\nonumber\\  
   && \sqrt{(2j+1)(2j'+1)}  
    \begin{Bmatrix}
  j & \lambda & I \\
  I & 1 & j' \
  \end{Bmatrix}  \begin{Bmatrix}
  l & 1/2 & j \\
  1 & j' & 1/2 
  \end{Bmatrix} \nonumber\\
&&\sqrt{6}\,(-u_ju_j'+v_jv_j') \delta(n_\lambda:n_\lambda')\delta(\lambda,\lambda') .
\end{eqnarray}

In this expression  $n$ and $n'$ are the number of phonons of each the initial and final configurations, respectively,
and the radial and orbital quantum numbers of the involved quasiparticle states $j$ and $j'$ should be equal.


\section{Results}\label{results}

We have calculated the expected signal for a xenon detector and for a germanium detector, considering neutralino masses between 10 and 100 GeV, by applying the formalism of Section \ref{sec:rates}. For the calculation of the rates, we have taken into account the limits imposed on the cross-section mass plane \cite{PDG:2020} given by the Xenon1T exclusion limit \cite{Xenon1T:2019}.

To start with, and following the steps discussed in the previous sections we have calculated the spectrum of the odd-mass 
$^{73}$Ge and $^{131}$Xe nuclei. This is done in two steps, namely:

\subsection{Nuclear structure of the even-mass mother nuclei:}
{ {
The calculation of the wave function of the low-lying quadrupole excitations in the even-even mother nuclei, is needed in order to determine the strength functions and coupling coefficients to be used in the calculation of the odd mass isotopes.  
To verify the accuracy of the description, which is based in the use of the Quasiparticle Random Phase Approximation, we have calculated E2 matrix elements and B(E2) transition probabilities and compared the results with the available data and other theoretical calculations.}}

{ {The one-phonon energies, of the even mass nuclei $^{72}$Ge and $^{130}$Xe, were fixed at the experimentally determined values of 0.691 MeV ($0^+_1$) and 0.834 MeV ($2^+_1$) in  $^{72}$Ge and 0.536 MeV ($2^+_1$) in $^{130}$Xe, respectively.
The comparison between the theoretical and experimental energy levels in $^{130}$Xe is shown in Table \ref{levelsxe}.
It is then seen that the agreement between theory and data is quite good, although the QRPA formalism assumes 
a purely harmonic response to nuclear correlations. for the case of $^{72}$Ge the low-lying monopole and quadrupole excitations have been adjusted in order to reproduce the experimental values, as said before. States of larger multi-polarities, for the A=130 case,  are lying at higher energies, around 1.8 MeV to 2.5 MeV and their mixing with quasiparticle states becomes negligible in dealing with the low-lying states of the odd-mass nuclei, whose wave functions are dominated by the coupling to the first excited monopole and quadrupole states (see next subsection).}}
{ {
\begin{table}[h!]
\centering
\begin{tabular}{ccccc}
\hline
State($J^{\pi}$ & Energy(Theory) MeV & Energy(Experiment)MeV\\
\hline
0$^+$(g.s.)& 0.0&0.0 \\
2$^+$(adjusted)& 0.536&0.536 \\
4$^+$& 1.250&1.150 \\
2$^+$&1.320&1.200 \\
\hline
\end{tabular}
\caption{Calculated (QRPA) and experimental (Ref.\cite{Morrison:2020}) values of low-lying  one phonon states in $^{130}$Xe..}
\label{levelsxe}
\end{table}
}}

{ {In order to give an idea about the quality of the calculations, we are quoting the B(E2) values listed in Table III of Ref.\cite{Morrison:2020} and compared them with our results for 
the case $^{130}$Xe (see Table \ref{evenmass})}}.
{ {
\begin{table}[h!]
\centering
\begin{tabular}{ccccc}
\hline
Transition($I_i$ $ \rightarrow$ $I_f$) & B(E2)(W.u.) & Source\\
\hline
$2^+$(first exc.) $ \rightarrow $ $0^+$(g.s) & 32(3) & (experimental value)(a)\\
                         & 21 ($e_n$=0.5e,$e_p$=1.5e) &  (b)\\
                         & 20  ($e_n$=0.5e,$e_p$=1.5e) & (c)\\
                         & 18.10 ($e_n$=0.5e,$e_p$=1.5e)  &  (present value)  (d)\\
                         & 22.84 ($e_n$=0.84e,$e_p$=1.68e)  & (present value) (e)\\
\hline
\end{tabular}
\caption{Experimental and calculated value for the B(E2) transition from the low-lying quadrupole excitation to 
the ground state  of $^{130}$Xe. The values have been taken from the work of Ref.\cite{Morrison:2020}, and they are the
experimental value (a), and the theoretical results (b) and (c) obtained by using shell model interactions . The values indicated by (d) and (e) are the results of the the present QRPA calculations}
\label{evenmass}
\end{table}
}}

{ {Finally, about the theoretical results corresponding to the even-mass cores, we shall compare the calculated values of the couplings $\chi_{\lambda}$ with their mass-dependent estimates given in Ref.\cite{Bohr-Mottelson:1974}.
The values are given in Table \ref{chilambda}.

\begin{table}[h!]
\centering
\begin{tabular}{ccccc}
\hline
Mass & Multipole-parity & Calculated(QRPA) & A-dependent value Ref.\cite{Bohr-Mottelson:1974}\\
\hline
72&  2(+)  & 0.00568 & 0.00782  \\
130& 2(+) &  0.00291 & 0.00191    \\
\hline
\end{tabular}
\caption{Calculated (QRPA) and mass-dependent estimates of the coupling strength $\chi_{\lambda}$, for quadrupole one-phonon states of the even mass cores. The values are given in units of MeV/fm$^{4}$}
\label{chilambda}
\end{table}
}}

\newpage
\subsection{Nuclear structure of the odd-mass nuclei:}
{ {
As mentioned before, the choice of the odd mass Ge and Xe nuclei is basically related to the importance of the mother nuclei 
in experiments related to double beta decay processes, as it is well documented in the literature.
Here, we have assumed that both of the even mass mother nuclei are spherical, although this assumption for the case of Ge may be taken with care due to deformations \cite{Wong:2022,Pritychenko:2022}. This issue was investigated in the past in great detail  by applying model descriptions based on the interaction between valence particles in the so-called Alaga Model \cite{Almar:1972}}}.

\subsubsection{Single particle estimates of the spin matrix elements}

{ {As stated in the previous section, the differential cross section for the elastics scattering of WIMPs by nuclei depends
solely on the expectation value of the spin between the involved nuclear states. Therefore we shall pay attention to the hindrance and fragmentation effects induced by both the pairing correlations and the coupling to the phonons of the even-mass core and before showing the results of the spin-matrix elements between states of one-quasiparticle coupled to monopole or quadrupole one  phonon states, and in order to have an idea about the effects of the coupling, we shall show the results for the matrix element  of the spin operator between pure neutron single particle states. They are given in Table ~\ref{spin_sp}.

\begin{table}[h!]
\centering
\begin{tabular}{ccccc}
\hline
final state (Nlj) & initial state (Nlj) &$\langle S \rangle$\\
\hline
4d3/2&4d3/2&-1.549 \\
4d3/2&4d5/2&3.098\\
4s1/2&4s1/2&2.449\\
4d5/2&4d5/2&2.898\\
4g7/2&4g7/2&-2.494\\
4g7/2&4g9/2&4.216\\
4g9/2&4g9/2&3.496\\
\hline
\end{tabular}
\caption{Calculated spin expectation values for single neutron states, $\langle S \rangle$, for active neutron states of positive parity in the shell with neutron number between 40 and 70 .}
\label{spin_sp}
\end{table}
}}

\newpage
\subsubsection*{Energy levels of the odd-mass nuclei}

{ {Here we shall present the results of the calculations of the low-lying energy levels of the odd mass nuclei $^{73}$Ge and $^{131}$Xe, which have been obtained by applying the one-phonon-one quasiparticle scheme.
 In both odd-mass nuclei the coupling to octupole states of the mother even mass nuclei is not taken into account since these states are lying at higher energies \cite{Wong:2022,Ahmed:2017}.
 
\begin{itemize}

\item{Results for $^{73}$Ge}

The case of $^{73}$Ge is somehow difficult to explain in a purely spherical basis, than the case of $^{131}$Xe (see next item). The coupling of quasi-neutrons to monopole and quadrupole phonons is very sensitive to the quasiparticle spectrum. What we have obtained, in support of the theoretical assumptions, is a quasi degeneracy between the first 
excited $J^{\pi}={5/2,7/2}^{+}$ and ${1/2}^{-}$ states which have energies of the order of few tens of keV above the $J^{\pi}={9/2}^{+}$ ground state. The results are shown in the next Table.\ref{oddspectrumge}

\begin{table}[h!]
\centering
\begin{tabular}{ccccc}
\hline
       &Energy (MeV) &  & \\ 
\hline
$j(\pi)$& Exp & Theory\\
9/2(+)&g.s&0.0\\
5/2(+)&0.013&0.041\\
1/2(-)&0.066&0.096\\
7/2(+)&0.068&0.110\\
\hline

\end{tabular}

\caption{Energy levels of low-lying states in $^{73}$Ge. Experimental values are compared to the results obtained by applying the quasiparticle-phonon coupling model}\label{oddspectrumge}
\end{table}

\item{Results for $^{131}$Xe }

As done for the case of the even mass nucleus $^{130}$Xe, and in order to get an idea about the validity of the quasiparticle-phonon coupling scheme for the odd-mass nucleus $^{131}$Xe, we show in Table \ref{oddspectrum} the comparison between the experimental and the calculated values of the energies and spin assignments for the low-lying states.
As it is seen from the results the agreement is quite good, except for the negative parity  state with $j^\pi$=${9/2}^-$ tentatively identify at about 350 keV, which in the quasiparticle phonon coupling scheme is predicted with a shift of the order of 500 keV respect to the $j^\pi$=${11/2}^-$ state, both members of the $h_{11/2}$ multiplet. Notice that the shift is just of the order of the one phonon energy.

\begin{table}[h!]
\centering
\begin{tabular}{ccccc}
\hline
       &Energy (MeV) &  & \\ 
\hline
$j(\pi)$& Exp & Theory\\
3/2(+)&g.s&0.0\\
1/2(+)&0.090&0.098\\
11/2(-)&0.170&0.130\\
9/2(-)&0.350&0.620\\
5/2(+)&0.570&0.534\\
3/2(+)&0.570&0.560\\
\hline

\end{tabular}

\caption{Energy levels of low-lying states in $^{131}$Xe. Experimental values are compared to the results obtained by applying the quasiparticle-phonon coupling model}\label{oddspectrum}
\end{table}

\end{itemize}
}}

{ {Then, from the analysis of the results provided by the QRPA method for the even mass nuclei and for the quasiparticle -phonon 
coupling scheme, one may conclude by saying that the results of the nuclear structure part of the calculations are rather acceptable when compared to data, for $^{130}$Xe and $^{131}$Xe. The case of $^{73}$Ge is a bit difficult, because of deformation effects in the even mass core $^{72}$Ge.
Nevertheless the small splitting between the ground state $(9/2)+$ and the first excited $(1/2)-$ state (which is a pure $p(1/2)$ state ) is well reproduced. }}

Next, we shall proceed with the calculation of the expectation value of the spin operator, which is the main purpose of the present work. We have calculated the spin expectation values for the $^{73}$Ge and $^{131}$Xe target nuclei.

\newpage

\subsection{Expectation value of the spin operator:}

The calculated amplitudes of the wave function of the ground state of each of the odd-mass nuclei $^{73}$Ge (I$^\pi$=9/2$^+$)  and $^{131}$Xe (I$^\pi$=3/2$^+$)
are written 
\begin{eqnarray}
|^{73}Ge (I^\pi=9/2^+)_{g.s}\rangle&=&0.9816|4g_{9/2}\rangle-0.0946|4g_{7/2}\otimes 2^+\rangle  \nonumber \\
&-&0.1291|4g_{9/2}\otimes 0^+\rangle -0.1046|4g_{9/2}\otimes 2^+\rangle \nonumber \\
|^{131}Xe (I^\pi=3/2^+)_{g.s}\rangle&=&0.9950|4d_{3/2}\rangle + 0.0897|4d_{3/2}\otimes 2^+\rangle \nonumber \\
&+&0.0619|4d_{5/2}\otimes 2^+\rangle .
\end{eqnarray}
{ {The single particle orbits are denoted in the (Nlj) notation, where N is the principal quantum number of the harmonic oscillator orbit, l is the orbital quantum number and j is the angular momentum}}.

{ { The matrix elements of the spin operator, whose expression for the quasiparticle-phonon coupling scheme .is given already (See Eq.(\ref{matrix_elements}), are the main element entering the calculation of the the differential cross section for the neutralino-nucleus elastic scattering of Eq.\ref{sigma_SD}. The orbital part does not participate explicitly because we are dealing with quasi-neutrons}}

In Table~\ref{spin_exp} we summarize our results for the expectation value of the spin operator in the neutron channel and compare them to the results of previous calculations quoted in Refs.\cite{Engel:1992,Pirinen:2019,Dimitrov:1994}. The results for the proton channel are negligible because the cancellation induced by their occupation numbers in the BCS approach.

\begin{table}[h!]
\centering
\begin{tabular}{ccccc}
\hline
nucleus & J$\pi$ & Ref. &$\langle S_n \rangle$& $\langle S_p \rangle$\\
\hline
$^{131}$Xe& $3/2+$&This work &-0.1349&0.0\\
& &QTDA Engel et al. \cite{Engel:1992}&  -0.236 &  -0.041\\
&&IFBM-2 Pirinen et al. \cite{Pirinen:2019}  & -0.188 &-0.0222\\
\hline
$^{73}$Ge & $9/2+$ & This work &0.2871 &0.0\\
& & Hybrid Dimitrov et al. \cite{Dimitrov:1994} &  0.378 &  0.03\\
& & CEFT Klos et al. \cite{Klos:2013}  &  0.439 &  0.031\\
\hline
\end{tabular}
\caption{Calculated spin expectation values for neutrons $\langle S_n \rangle$ and protons $\langle S_p \rangle$ of $^{131}$Xe and $^{73}$Ge in different models. QTDA: Quasi Tamm-Dancoff Approximation \cite{Engel:1992},  IFBM-2:  microscopic interacting boson-fermion model \cite{Pirinen:2019}, CEFT: Chiral Effective Field Theory \cite{Klos:2013}.}
\label{spin_exp}
\end{table}

{ {As it is seen from these results the inclusion of backward going amplitudes, that is by the way of the QRPA formalism, reduces the value of $\langle S_n \rangle$ making it closer to the value calculated by Pirinen et al. \cite{Pirinen:2019}, for $^{131}$Xe,  but definitively smaller, for the case of $^{73}$Ge, than the value given by Dimitrov et al. \cite{Dimitrov:1994} in their Hybrid model, and that of Klos et al.\cite{Klos:2013} obtained using the Chiral Effective Field Theory (CEFT) model.}}

{ {As a general comment about the order of magnitude of the matrix elements of the spin operator due to coupling effects between quasiparticles and phonons, it can be said that this mechanism reduces the matrix elements of Table \ref{spin_sp} by factors of the order of 10 or larger, as shown by the values listed in Table \ref{spin_exp}}}.

\subsection{WIMPs-nucleus cross section}
Next, we have calculated the cross section given by Eq.(\ref{sigma_SD}) for different values of the mass of the WIMPs. The results are given in Table~\ref{sigma_masas}.

\begin{table}[h!]
\centering
\begin{tabular}{cccc}
\hline
nucleus & J &m$_\chi$[GeV]&$\sigma_{SD}$ [cm$^2$] \\
\hline
$^{131}$Xe& $3/2+$&10&1.60$\times 10^{-45}$ \\
& &50& 1.21$\times 10^{-46}$\\
& &100& 1.79$\times 10^{-46}$\\
\hline
$^{73}$Ge& $9/2+$&10& 4.75$\times 10^{-45}$\\
& &50& 2.66$\times 10^{-46}$\\
& &100& 5.04$\times 10^{-46}$\\
\hline
\end{tabular}
\caption{Calculated cross-sections following Eq. \ref{sigma_SD} for different WIMP's masses using our nuclear model.}
\label{sigma_masas}
\end{table}

The results of Table~\ref{sigma_masas}, for the detection of a fermionic cold-dark-matter particle such as the neutralino, 
have been obtained by fixing the SUSY parameters entering the equations introduced in Section \ref{dmmodel}.
The s-quark-mass and the Higgs pseudo-scalar mass were fixed at the values 1500 GeV and 500 GeV, respectively \cite{Pdg:2019}. 
We fixed also the parameter $\tan{\beta}=10$ \cite{Ellis:2000,Cerdeno:2001,Murakami:2001}, and varied the parameter $\mu$. The value of the  parameter $M$ was determined as a function of $\mu$ and $m_{\chi}$.

\begin{figure}
\begin{center}
\includegraphics[width=1.0\textwidth]{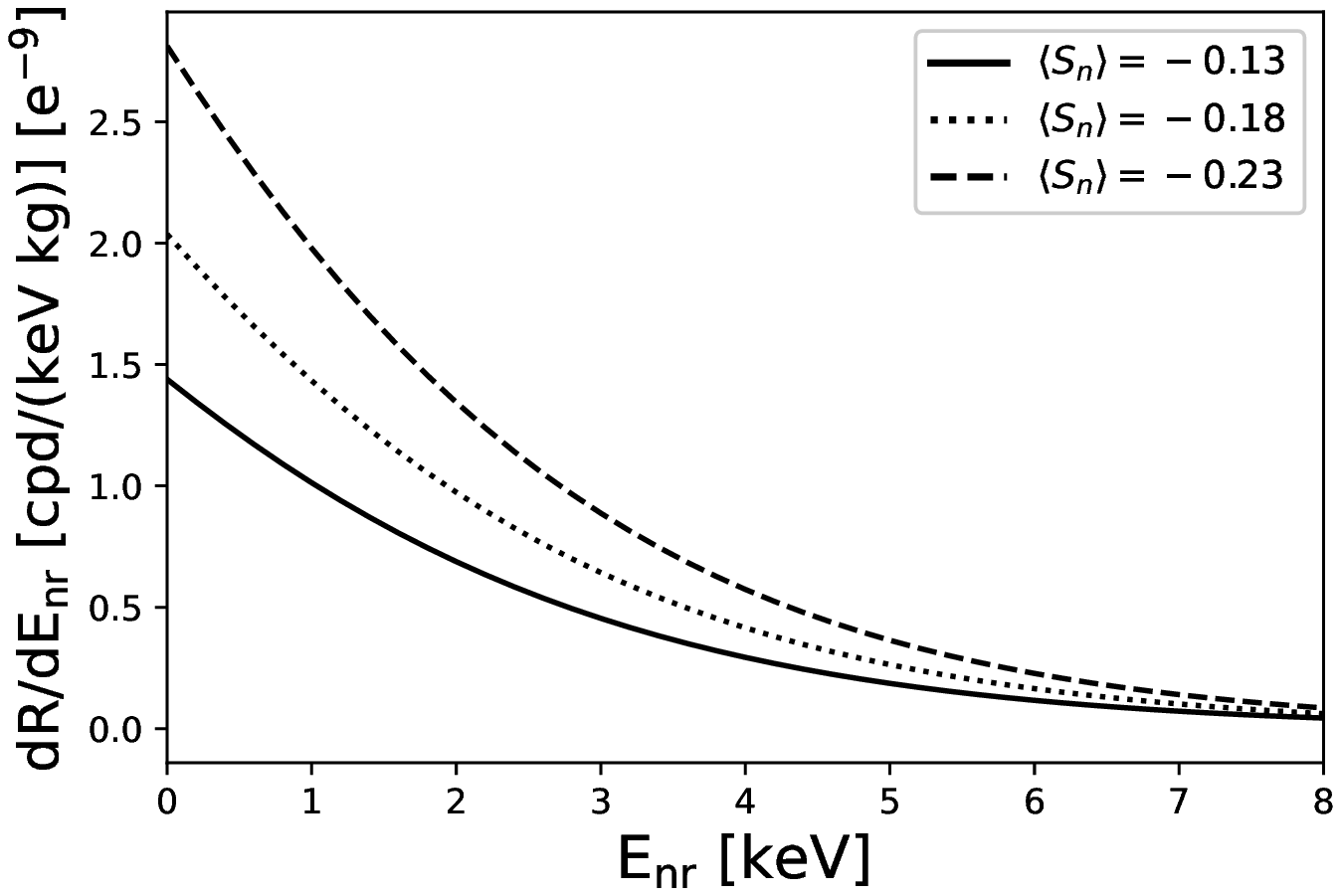}
\end{center}
\caption{Results for $m_{\chi}=10$ GeV for $^{73}$Ge. Upper panel: differential recoil rate $dR/dE_{nr}$ as a function of the nuclear recoil energy $E_{nr}$.  
Solid line: spin expectation value (this work); dashed line: spin expectation value from Ref. \cite{Klos:2013}, dotted line spin expectation value from Ref. \cite{Dimitrov:1994}.}\label{gem10}
\end{figure}

\begin{figure}
\includegraphics[width=1.0\textwidth]{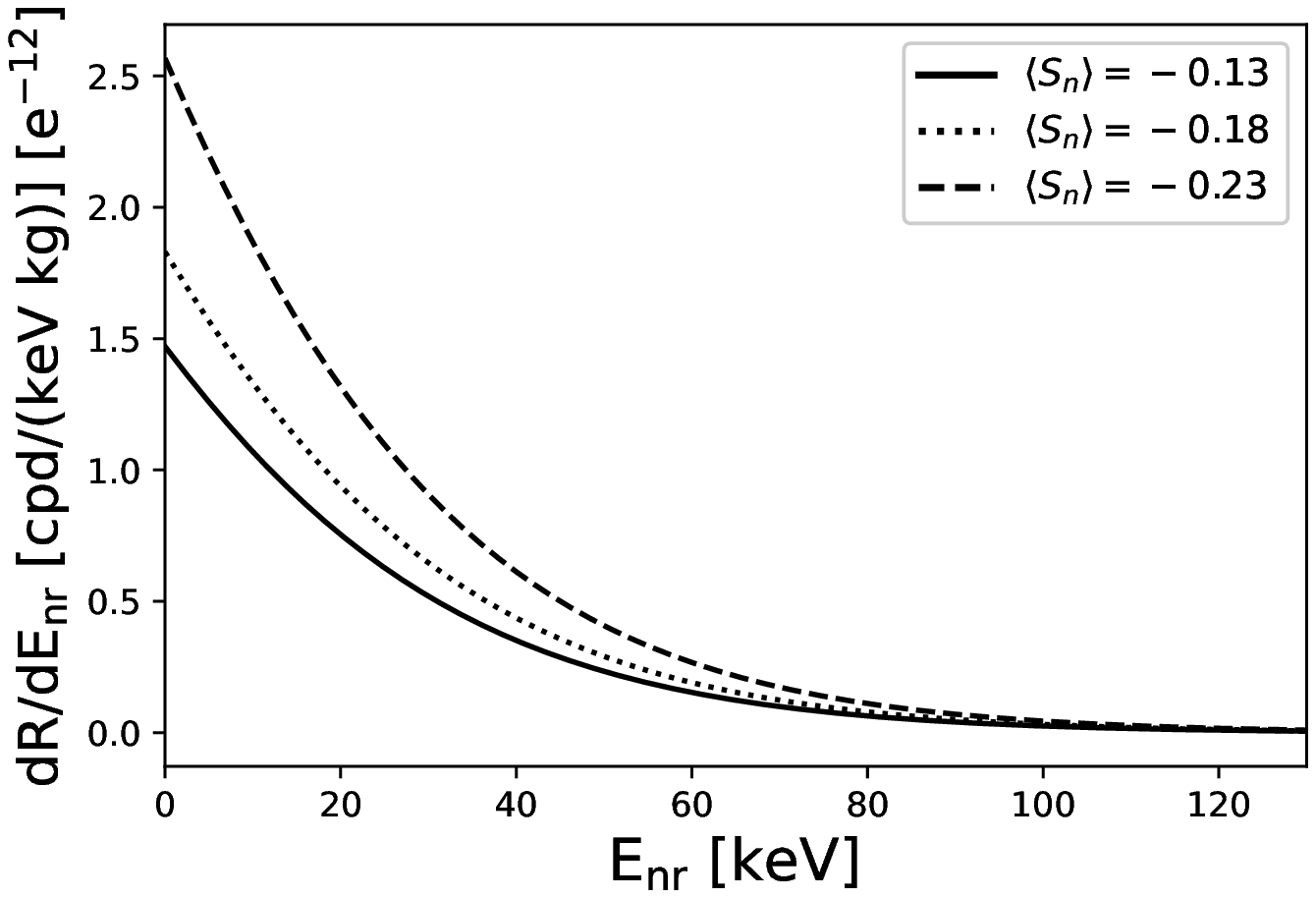}\label{gem50}
\caption{Same as Figure \ref{gem10} but for $m_{\chi}=50$ GeV.}
\end{figure}

\begin{figure}
\includegraphics[width=1.0\textwidth]{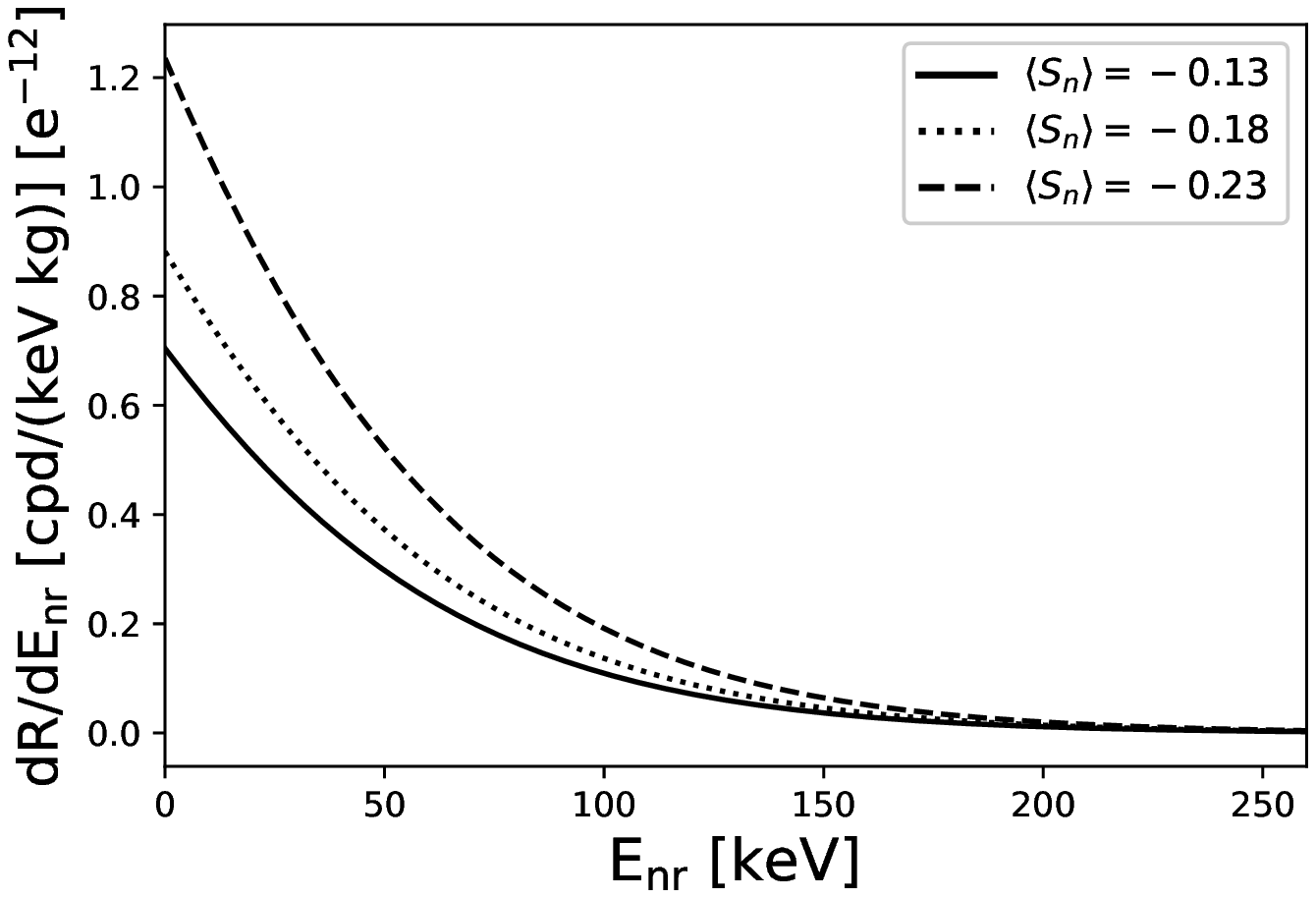}\label{gem100}
\caption{Same as Figure \ref{gem10} but for $m_{\chi}=100$ GeV.}
\end{figure}

\begin{figure}
\begin{center}
\includegraphics[width=1.0\textwidth]{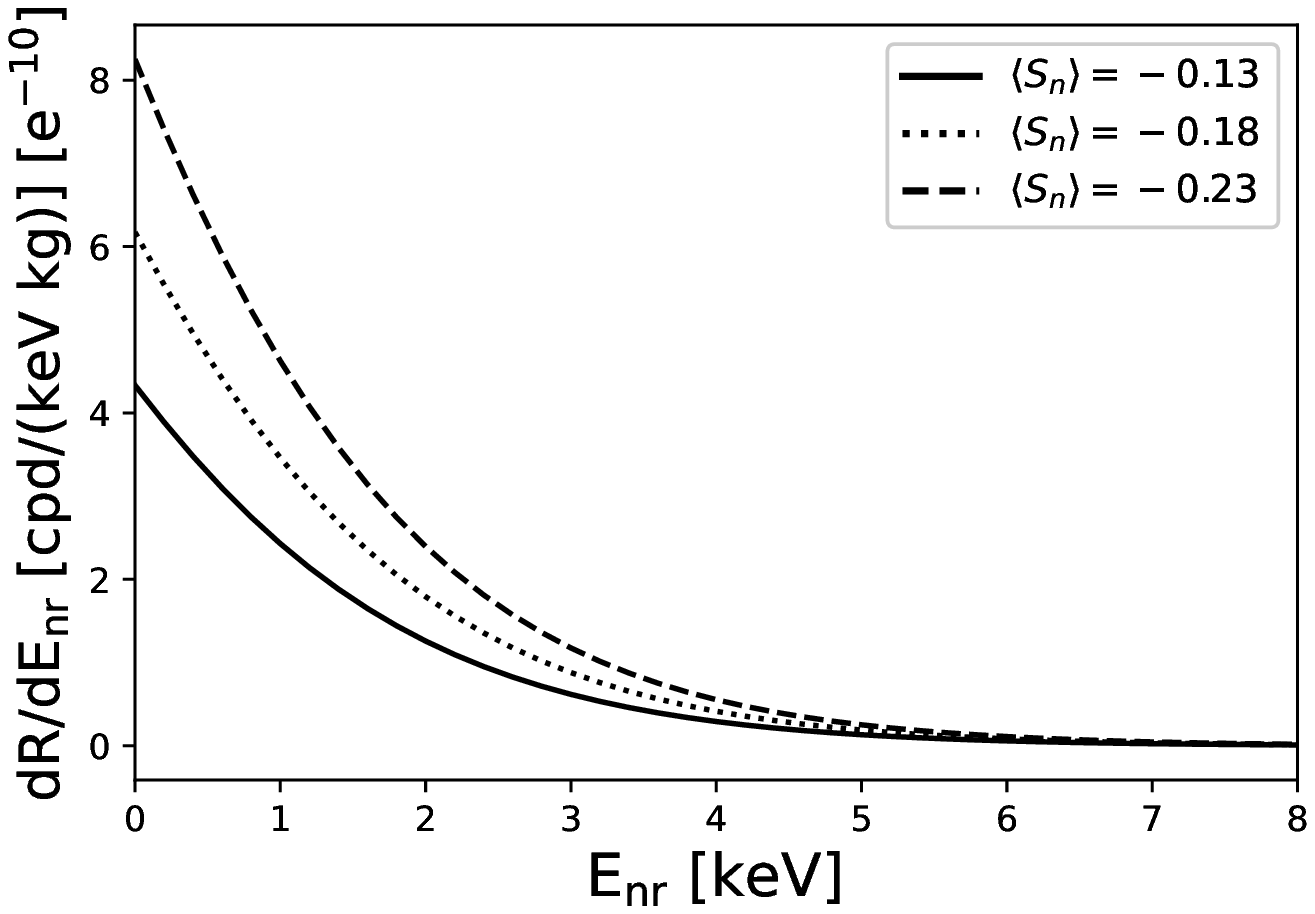}
\end{center}
\caption{Results for $m_{\chi}=10$ GeV for $^{131}$Xe. Upper panel: differential recoil rate $dR/dE_{nr}$ as a function of the nuclear recoil energy $E_{nr}$. 
Solid line: spin expectation value (this work); dashed line: spin expectation value from Ref. \cite{Engel:1992}, dotted line spin expectation value from Ref.\cite{Pirinen:2019}.}
\label{xem10}
\end{figure}

\begin{figure}
\begin{center}
\includegraphics[width=1.0\textwidth]{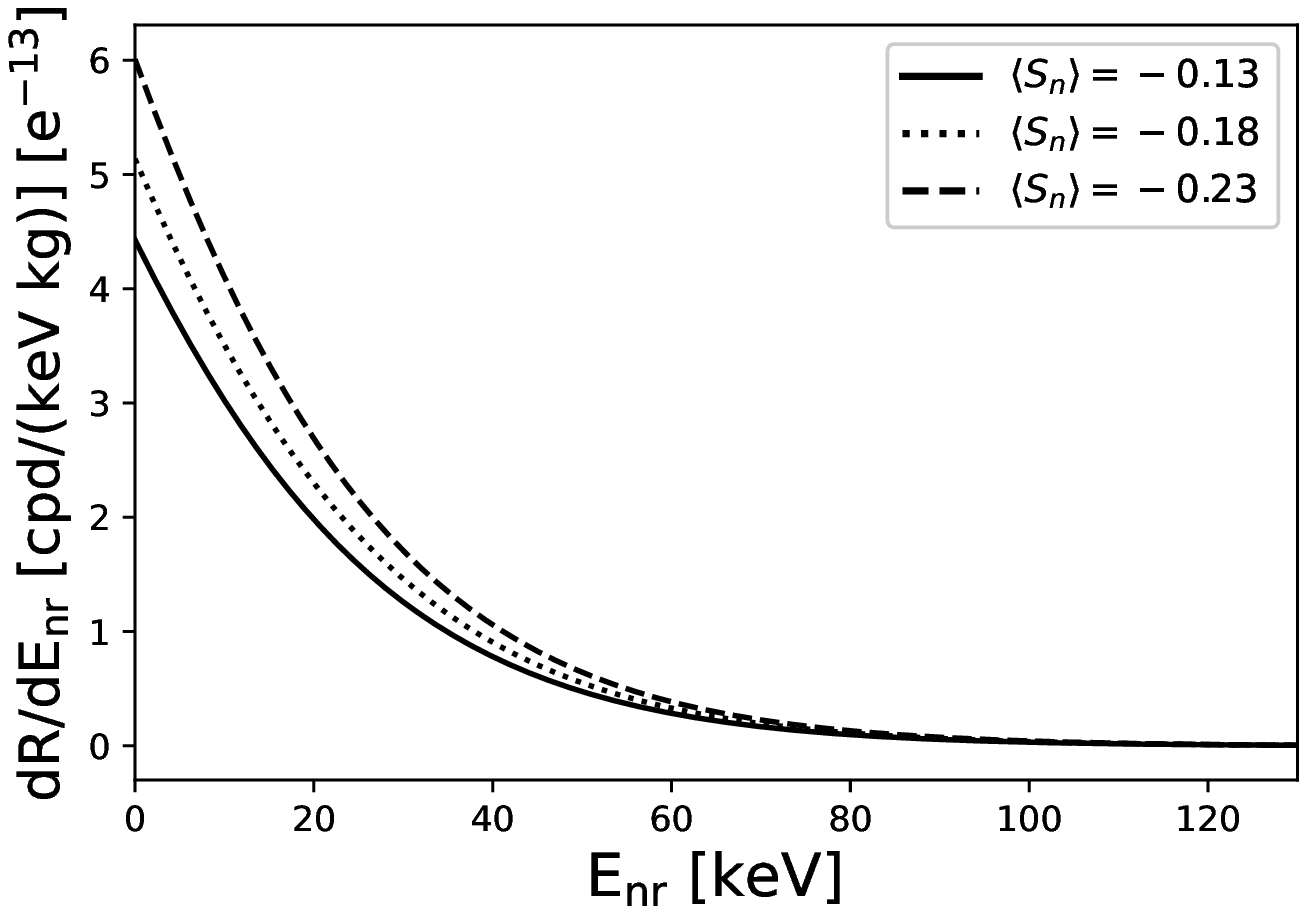}
\end{center}
\caption{Same as Figure \ref{xem10} but for $m_{\chi}=50$ GeV.}
\label{xem50}
\end{figure}

\begin{figure}
\begin{center}
\includegraphics[width=1.0\textwidth]{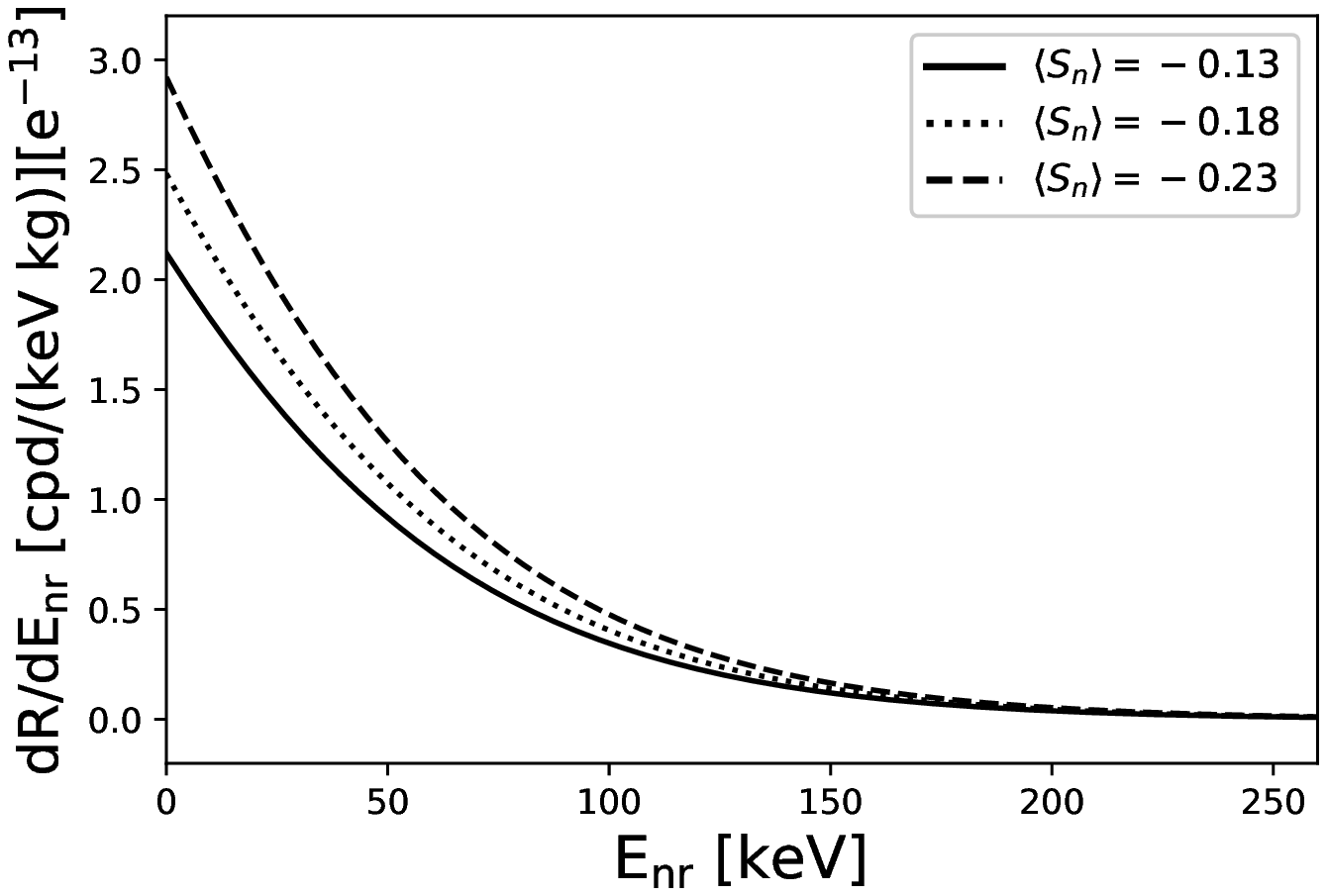}
\end{center}
\caption{Same as Figure \ref{xem10} but for $m_{\chi}=100$ GeV.}
\label{xem100}
\end{figure}

Results for the target nucleus $^{73}$Ge are shown in Figs.\ref{gem10}-\ref{gem100}, respectively, while those 
for the $^{131}$Xe detector are shown in Figs.\ref{xem10}-\ref{xem100}.
In the upper panel of the aforementioned figures we show the spin-dependent differential direct detection rate (see Eq. \ref{rate2}) as a function of the nuclear recoil energy for different WIMP's masses and for June 2, where the laboratory speed is maximum. Solid-lines represent the results obtained with our spin expectation value, dashed-lines are the ones corresponding to Ref. \cite{Engel:1992}, and dotted-lines are those taken from Ref. \cite{Pirinen:2019}. 

In Table \ref{gammas} we are listing the ratio ($\Gamma$) between the recoil rates obtained with different spin expectation values. From the behaviour of $\Gamma$ we note that, for lower WIMP masses, there is a higher sensitivity upon the spin expectation value making relevant the choice of the nuclear model to be used.  Different values of the spin content can amplify the rates by up to 2.5 times. The results are independent of the approximation used for the transfer of momentum.

\begin{table}[h!]
\centering
\begin{tabular}{cccc}
\hline
nucleus & m$_\chi$[GeV]&ratio(a)&ratio(b) \\
\hline
$^{73}$Ge & 10  & 0.512 & 0.707\\
          & 50  & 0.573 & 0.803\\
          & 100 & 0.570 & 0.799 \\
\hline
$^{131}$Xe    & 10  & 0.525 & 0.702\\
          & 50  & 0.737 & 0.862\\
          & 100 & 0.726 & 0.854 \\
\hline
\end{tabular}
\caption{Ratios of the counting rates  for the calculated spin expectation values, as a function of the WIMPs mass  m$_\chi$.
The columns denoted as ratio (a) and ratio (b) correspond to: (a)  $\frac{dR/dE_{nr}|_{\langle S_n \rangle=0.43}}{dR/dE_{nr}|_{\langle S_n \rangle=0.28}}$, and  (b) $\frac{dR/dE_{nr}|_{\langle S_n \rangle=0.37}}{dR/dE_{nr}|_{\langle S_n \rangle =0.28}}$ , for the case of $^{73}$Ge, respectively, and (a) $\frac{dR/dE_{nr}|_{\langle S_n \rangle=-0.23}}{dR/dE_{nr}|_{\langle S_n \rangle=-0.13}}$,  and (b) $\frac{dR/dE_{nr}|_{\langle S_n \rangle=-0.18}}{dR/dE_{nr}|_{\langle S_n \rangle =-0.13}}$ for $^{131}$Xe.}
\label{gammas}
\end{table}

\section{Conclusions}\label{conclusions}
In this work, we have studied the nuclear response in dark matter direct detection experiments. Particularly, we have focus the attention on odd-mass germanium and xenon
isotopes as detectors. We have described dark-matter particles within the Minimal Supersymmetric Standard Model  and studied the effects due to different masses and SUSY parameters within the limits established by the latest results of the Xenon1T collaboration. We have computed the cross-sections and differential rates for the elastic scattering of WIMPs by nuclei adopting the quasiparticle-vibration coupling model for the description of the nuclear structure of the odd-mass nuclei 
$^{73}$Ge and $^{131}$Xe.
{ { The results for the nuclear structure components of the calculations, both for the even-mass cores and for the odd-mass isotopes show that the quasiparticle-phonon picture provides a reasonable description of their low-energy properties, namely: the energy levels, both for the even and odd mass nuclei, the B(E2) values for transitions in the even mass nuclei, the occupation factors and quasiparticle energies and angular momentum sequences for the odd-mass nuclei.

We have obtained spin expectation values which are smaller than those reported in the literature until now. The differential rates obtained using our results are smaller than those obtained with other spin contents. Germanium was more sensitive to the changes in the spin content than Xenon.
We also found that there is a strong sensitivity, with respect to the nuclear spin expectation value, for light WIMPs masses. 
In contrast, when working with heavy mass neutralinos, the dependence with the transfer of momentum becomes relevant. 
{ {However, in spite of these details, the general trend of the results of the QRPA and quasiparticle-phonon coupling model agree quite satisfactory with those obtained with the QTDA of Engel et al.\cite{Engel:1992} and more closely to the results provided by the use of the IFBM (Interacting Fermion Boson Model) of Pirinen et al. \cite{Pirinen:2019}.}}
The changes in the recoil rate due to the zero momentum transfer approximation generates differences of about 30 percent with respect to the case of finite momentum case, especially at high energies. }}

\section*{Acknowledgments}
This work has been partially supported by the National Research Council of Argentina (CONICET) by the grant PIP 11220200102081CO, and by the Agencia Nacional de Promoci\'on Cient\'ifica y Tecnol\'ogica (ANPCYT) PICT 140492. O.C and T.T are members of the Scientific Research Career of the CONICET. Discussions with Dr M. E.Mosquera are gratefully acknowledged.

\bibliography{biblio.bib}

\providecommand{\noopsort}[1]{}\providecommand{\singleletter}[1]{#1}%
\begin{thebibliography}{10}

\bibitem{Rubin:1970}
V.~C. {Rubin} and J.~{Ford}, W.~Kent, ``{Rotation of the Andromeda Nebula from
  a Spectroscopic Survey of Emission Regions},'' {\em Astrophys. J.}, vol.~159,
  p.~379, 1970.

\bibitem{zwicky:1933}
F.~Zwicky, ``Die rotverschiebung von extragalaktischen nebeln,'' {\em Helv.
  Phys. Acta}, vol.~6, p.~110, 1933.

\bibitem{Clowe:2006}
D.~Clowe {\em et~al.}, ``A direct empirical proof of the existence of dark
  matter,'' {\em Astrophys. J.}, vol.~648, p.~L109, 2006.

\bibitem{Planck:2018}
N.~{Aghanim} {\em et~al.}, ``{Planck 2018 results. VI. Cosmological
  parameters},'' {\em Astronomy and Astrophysic}, vol.~641, p.~A6, 2020.

\bibitem{Schumann:2019}
M.~Schumann, ``Direct detection of {WIMP} dark matter: concepts and status,''
  {\em JPG}, vol.~46, p.~103003, 2019.

\bibitem{majumdar:2014}
D.~Majumdar, {\em Dark Matter: An Introduction}.
\newblock Taylor \& Francis, 2014.

\bibitem{Peccei:1977}
R.~D. Peccei and H.~R. Quinn, ``$\mathrm{CP}$ conservation in the presence of
  pseudoparticles,'' {\em Phys. Rev. Lett.}, vol.~38, pp.~1440--1443, Jun 1977.

\bibitem{chadhaday:2021}
F.~Chadha-Day, J.~Ellis, and D.~J.~E. Marsh, ``Axion dark matter: What is it
  and why now?,'' 2021.

\bibitem{Gelmini:2017}
G.~B. {Gelmini}, ``{Light weakly interacting massive particles},'' {\em Rept.
  Prog. Phys.}, vol.~80, p.~082201, 2017.

\bibitem{Goodman:1985}
M.~W. {Goodman} and E.~{Witten}, ``{Detectability of certain dark-matter
  candidates},'' {\em Phys. Rev. D}, vol.~31, p.~3059, 1985.

\bibitem{Wasserman:1986}
I.~Wasserman, ``{Possibility of Detecting Heavy Neutral Fermions in the
  Galaxy},'' {\em Phys. Rev. D}, vol.~33, p.~2071, 1986.

\bibitem{XENON1T:2018}
E.~{Aprile} {\em et~al.}, ``{Dark Matter Search Results from a One Ton-Year
  Exposure of XENON1T},'' {\em Phys. Rev. Lett.}, vol.~121, p.~111302, 2018.

\bibitem{LUX:2017}
D.~S. Akerib {\em et~al.}, ``{Limits on spin-dependent WIMP-nucleon cross
  section obtained from the complete LUX exposure},'' {\em Phys. Rev. Lett.},
  vol.~118, p.~251302, 2017.

\bibitem{EDELWEISS:2020}
Q.~Arnaud {\em et~al.}, ``{First germanium-based constraints on sub-MeV Dark
  Matter with the EDELWEISS experiment},'' {\em Phys. Rev. Lett.}, vol.~125,
  no.~14, p.~141301, 2020.

\bibitem{CoGeNT:2012}
C.~E. Aalseth {\em et~al.}, ``{CoGeNT: A Search for Low-Mass Dark Matter using
  p-type Point Contact Germanium Detectors},'' {\em Phys. Rev. D}, vol.~88,
  p.~012002, 2013.

\bibitem{Bednyakov:2007}
V.~A. {Bednyakov}, ``{Spin in the dark matter problem},'' {\em Physics of
  Particles and Nuclei}, vol.~38, pp.~326--363, May 2007.

\bibitem{Menendez:2012}
J.~{Men{\'e}ndez}, D.~{Gazit}, and A.~{Schwenk}, ``{Spin-dependent WIMP
  scattering off nuclei},'' {\em physical review d}, vol.~86, p.~103511, Nov.
  2012.

\bibitem{Marcos:2016}
C.~{Marcos}, M.~{Peir{\'o}}, and S.~{Robles}, ``{On the importance of direct
  detection combined limits for spin independent and spin dependent dark matter
  interactions},'' {\em JCAP}, vol.~2016, p.~019, 2016.

\bibitem{Bednyakov:2004}
V.~A. Bednyakov and F.~Simkovic, ``{Nuclear spin structure in dark matter
  search: The Zero momentum transfer limit},'' {\em Phys. Part. Nucl.},
  vol.~36, pp.~131--152, 2005.

\bibitem{Bednyakov:2006}
V.~A. {Bednyakov} and F.~{{\v{S}}imkovic}, ``{Nuclear spin structure in dark
  matter search: The finite momentum transfer limit},'' {\em Physics of
  Particles and Nuclei}, vol.~37, p.~S106, 2006.

\bibitem{Vietze:2014}
L.~Vietze, P.~Klos, J.~Men\'endez, W.~C. Haxton, and A.~Schwenk, ``{Nuclear
  structure aspects of spin-independent WIMP scattering off xenon},'' {\em
  Phys. Rev. D}, vol.~91, no.~4, p.~043520, 2015.

\bibitem{Bohr-Mottelson:1974}
A.~Bohr and B.~R. Mottelson, {\em Nuclear Structure}.
\newblock World Scientific Publishing Company, 1998.

\bibitem{Pirinen:2019}
P.~Pirinen, J.~Kotila, and J.~Suhonen, ``Spin-dependent wimp-nucleus scattering
  off 125te, 129xe, and 131xe in the microscopic interacting boson-fermion
  model,'' {\em Nuclear Physics A}, vol.~992, p.~121624, 2019.

\bibitem{Engel:1992}
J.~{Engel}, S.~{Pittel}, and P.~{Vogel}, ``{Nuclear Physics of Dark Matter
  Detection},'' {\em IJMP E}, vol.~1, p.~1, 1992.

\bibitem{Klos:2013}
P.~Klos, J.~Men\'endez, D.~Gazit, and A.~Schwenk, ``Large-scale nuclear
  structure calculations for spin-dependent wimp scattering with chiral
  effective field theory currents,'' {\em Phys. Rev. D}, vol.~88, p.~083516,
  Oct 2013.

\bibitem{Murakami:2001}
B.~{Murakami} and J.~D. {Wells}, ``{Nucleon scattering with Higgsino and W-ino
  cold dark matter},'' {\em Phys. Rev. D}, vol.~64, p.~015001, 2001.

\bibitem{Ellis:2000}
J.~{Ellis}, A.~{Ferstl}, and K.~A. {Olive}, ``{Re-evaluation of the elastic
  scattering of supersymmetric dark matter},'' {\em Phys. Lett. B}, vol.~481,
  p.~304, 2000.

\bibitem{Cerdeno:2001}
D.~Cerdeno, S.~Khalil, and C.~Munoz, ``{Large dark matter cross-sections from
  supergravity and superstrings},'' in {\em {5th International Conference on
  Particle Physics and the Early Universe}}, 2001.

\bibitem{Djouadi:2008}
A.~Djouadi, ``The anatomy of electroweak symmetry breaking tome ii: The higgs
  bosons in the minimal supersymmetric model,'' {\em Phys. Rep.}, vol.~459,
  p.~1, 2008.

\bibitem{Pdg:2019}
M.~o. Tanabashi, ``Review of particle physics,'' {\em Phys. Rev. D}, vol.~98,
  p.~030001, Aug 2018.

\bibitem{Fushimi:2020}
K.~{Fushimi}, M.~E. {Mosquera}, and O.~{Civitarese}, ``{MSSM WIMPs-Nucleon
  cross-section for E{\ensuremath{\chi}} less than 500GeV},'' {\em
  International Journal of Modern Physics E}, vol.~29, pp.~2050072--446, Jan.
  2020.

\bibitem{Freese:2012}
K.~Freese, M.~Lisanti, and C.~Savage, ``{Colloquium: Annual modulation of dark
  matter},'' {\em Rev. Mod. Phys.}, vol.~85, p.~1561, 2013.

\bibitem{Freese:1988}
K.~Freese, J.~Frieman, and A.~Gould, ``Signal modulation in cold-dark-matter
  detection,'' {\em Phys. Rev. D}, vol.~37, p.~3388, 1988.

\bibitem{Jungman:1996}
G.~Jungman, M.~Kamionkowski, and K.~Griest, ``{Supersymmetric dark matter},''
  {\em Phys. Rept.}, vol.~267, p.~195, 1996.

\bibitem{Gelmini:2015}
G.~B. Gelmini, ``{TASI 2014 Lectures: The Hunt for Dark Matter},'' in {\em {}},
  2015.

\bibitem{Sahu:2016}
R.~Sahu and V.~K.~B. Kota, ``{Deformed Shell Model Study of Heavy $N = Z$
  Nuclei and Dark Matter Detection},'' {\em Nucl. Theor.}, vol.~35, pp.~22--33,
  2016.

\bibitem{Weinberg}
S.~Weinberg, {\em The Quantum Theory of Fields}, vol.~1.
\newblock Cambridge University Press, 1995.

\bibitem{Tovey:2000}
D.~R. Tovey, R.~J. Gaitskell, P.~Gondolo, Y.~A. Ramachers, and L.~Roszkowski,
  ``{A New model independent method for extracting spin dependent
  (cross-section) limits from dark matter searches},'' {\em Phys. Lett. B},
  vol.~488, pp.~17--26, 2000.

\bibitem{Dimitrov:1994}
V.~Dimitrov, J.~Engel, and S.~Pittel, ``{Scattering of weakly interacting
  massive particles from Ge-73},'' {\em Phys. Rev. D}, vol.~51, pp.~291--295,
  1995.

\bibitem{Ring-Schuck:1980}
P.~Ring and P.~Schuck, {\em The Nuclear Many-Body Problem}.
\newblock Physics and astronomy online library, Springer, 2004.

\bibitem{Civitarese:1998}
J.~{Suhonen} and O.~{Civitarese}, ``{Weak-interaction and nuclear-structure
  aspects of nuclear double beta decay},'' {\em Physics Reports}, vol.~300,
  pp.~123--214, Jan. 1998.

\bibitem{odd}
O.~{Civitarese}, R.~A. {Broglia}, and D.~R. {Bes}, ``{Role of the Pauli
  principle in the spectrum of $^{211}$Pb},'' {\em Physics Letters B}, vol.~72,
  pp.~45--48, Dec. 1977.

\bibitem{PDG:2020}
{Particle Data Group}, P.~A. {Zyla}, {\em et~al.}, ``{Review of Particle
  Physics},'' {\em Progress of Theoretical and Experimental Physics},
  vol.~2020, p.~083C01, 2020.

\bibitem{Xenon1T:2019}
E.~{Aprile} and {Xenon Collaboration}, ``{Light Dark Matter Search with
  Ionization Signals in XENON1T},'' {\em Phys. Rev. Lett.}, vol.~123,
  p.~251801, 2019.

\bibitem{Morrison:2020}
L.~{Morrison} {\em et~al.}, ``{Quadrupole deformation of 130Xe measured in a
  Coulomb-excitation experiment},'' {\em Physical Review C}, vol.~102,
  pp.~054304--1,054304--13, Nov. 2020.

\bibitem{Wong:2022}
C.~{Wong} {\em et~al.} {\em Physical Review C}, vol.~106, 2022.

\bibitem{Pritychenko:2022}
B.~{Pritychenko} {\em et~al.} {\em Nuclear Physics A}, vol.~1027, 2022.

\bibitem{Almar:1972}
R.~Almar, O.~Civitarese, F.~Krmpotic, and J.~Navaza, ``Structure of the odd
  mass ge isotopes with a particle phonon coupling.,'' {\em Physical Review C},
  vol.~C 6, 1972.

\bibitem{Ahmed:2017}
I.~M. {Ahmed} {\em AIP Proceedings}, vol.~1888, 2017.

\end{thebibliography}
\bibliographystyle{ieeetr}
\end{document}